\documentclass[pra,twocolumn,showpacs,superscriptaddress,floatfix]{revtex4}

\let\a=\alpha \let\b=\beta  \let\g=\gamma  \let\d=\delta \let\e=\varepsilon
\let\z=\zeta       
\let\m=\mu    \let\n=\nu    \let\x=\xi     \let\p=\pi    \let\r=\rho
\let\s=\sigma \let\t=\tau   \let\f=\varphi 
\let\ch=\chi  \let\ps=\psi   
 \let\D=\Delta  \let\Th=\Theta\let\L=\Lambda 
    \let\Si=\Sigma \let\F=\Phi    
\let\O=\Omega 

\font\tenmib=cmmib10\font\sevenmib=cmmib7\font\fivemib=cmmib5%
\mathchardef\Ba   = "050B  
\mathchardef\Bb   = "050C  
\mathchardef\Bg   = "050D  
\mathchardef\Bd   = "050E  
\mathchardef\Be   = "0522  
\mathchardef\Bee  = "050F  
\mathchardef\Bz   = "0510  
\mathchardef\Bh   = "0511  
\mathchardef\Bthh = "0512  
\mathchardef\Bth  = "0523  
\mathchardef\Bi   = "0513  
\mathchardef\Bk   = "0514  
\mathchardef\Bl   = "0515  
\mathchardef\Bm   = "0516  
\mathchardef\Bn   = "0517  
\mathchardef\Bx   = "0518  
\mathchardef\Bom  = "0530  
\mathchardef\Bp   = "0519  
\mathchardef\Br   = "0525  
\mathchardef\Bro  = "051A  
\mathchardef\Bs   = "051B  
\mathchardef\Bsi  = "0526  
\mathchardef\Bt   = "051C  
\mathchardef\Bu   = "051D  
\mathchardef\Bf   = "0527  
\mathchardef\Bff  = "051E  
\mathchardef\Bch  = "051F  
\mathchardef\Bps  = "0520  
\mathchardef\Bo   = "0521  
\mathchardef\Bome = "0524  
\mathchardef\BG   = "0500  
\mathchardef\BD   = "0501  
\mathchardef\BTh  = "0502  
\mathchardef\BL   = "0503  
\mathchardef\BX   = "0504  
\mathchardef\BP   = "0505  
\mathchardef\BS   = "0506  
\mathchardef\BU   = "0507  
\mathchardef\BF   = "0508  
\mathchardef\BPs  = "0509  
\mathchardef\BO   = "050A  
\mathchardef\BDpr = "0540  
\mathchardef\Bstl = "053F  

\newdimen\xshift \newdimen\xwidth \newdimen\yshift \newdimen\ywidth

\def\ins#1#2#3{\vbox to0pt{\kern-#2pt\hbox{\kern#1pt #3}\vss}\nointerlineskip}

\def\eqfig#1#2#3#4#5{
\par\xwidth=#1pt \xshift=\hsize \advance\xshift
by-\xwidth \divide\xshift by 2
\yshift=#2pt \divide\yshift by 2
{\hglue\xshift \vbox to #2pt{\vfil
#3 \includegraphics{#4.eps}
}\hfill\raise\yshift\hbox{#5}}}

\def\V#1{{\bf #1}}
\def\lis#1{{\overline#1}}

\font\titolo=cmbx12
\font\msytw=msbm10%
\def\RRR{\hbox{\msytw R}}
\def\tende#1{\,\vtop{\ialign{##\crcr\rightarrowfill\crcr
 \noalign{\kern-1pt\nointerlineskip} \hskip3.pt${\scriptstyle
 #1}$\hskip3.pt\crcr}}\,}
\def\EE{{\cal E}}\def\BB{{\cal B}}
\def\HH{{\cal H}}\def\NN{{\cal N}}
\def\defi{\,{\buildrel def\over=}\,}
\let\wt=\widetilde
\let\0=\noindent\def\*{\vskip2mm}

\def\AA{{\cal A}}\def\XX{{\cal X}}
\def\EE{{\cal E}}\def\BB{{\cal B}}
\def\HH{{\cal H}}\def\NN{{\cal N}}\def\CC{{\cal C}}

\font\msytw=msbm10
\def\RRR{\hbox{\msytw R}}

\def\ZZZ{\hbox{\msytw Z}}

\def\be{\begin{equation}}\def\ee{\end{equation}}
\def\beq{\begin{eqnarray}}\def\eeq{\end{eqnarray}}
\def\nn{\nonumber}

\let\0=\noindent
\usepackage{ifthen}
\usepackage{fancyhdr}
\def\iniz{\setcounter{equation}{0}
\rhead{\thepage}\lhead{{{{\small\thesection:}\ \SEC}}}}

\begin{document}

\centerline{\titolo
Thermodynamic limit for}
\centerline{\titolo isokinetic thermostats}
{\vskip3mm}

\centerline{G. Gallavotti${}^*$ and E. Presutti${}^@$}
\centerline{${}^*$ Fisica-INFN Roma1 and Rutgers U.}
\centerline{${}^@$ Matematica Roma2}
\centerline{\today}

{\vskip3mm}
\noindent {\bf Abstract}: {\it Thermostats models in space dimension
 $d=1,2,3$ for nonequilibrium statistical mechanics are considered and
 it is shown that, in the thermodynamic limit, the motions of
 frictionless thermostats and isokinetic thermostats coincide.}  {\vskip3mm}

\setcounter{equation}{0}
\setcounter{section}{0}
\def\SEC{Thermostats}
\section{Thermostats}\label{sec1}
\iniz

A {\it test system} of particles in a container $\O_0$ and $\n$
systems of particles in containers $\O_1,\ldots,\O_\n$ interact and
define a model of a system in interaction with $\n$ thermostats if the
particles in $\O_1,\ldots,\O_\n$ can be considered at fixed
temperatures $T_1,\ldots,T_\n$. 

A representation of the system is in Fig.1:

\eqfig{200}{86}{%
\ins{40}{12}{%
$x=(\V X_0,\dot{\V X}_0,\V X_1,\dot{\V X}_1,\ldots,\V X_\n,\dot{\V X}_\n)$}
}{fig1}{}

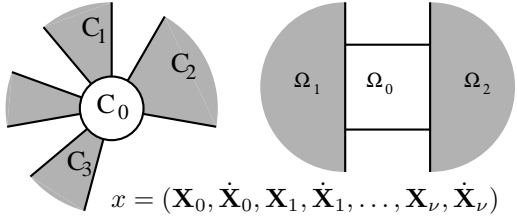
\noindent{Fig.1: \small\it The $1+\n$ finite boxes $\O_j\cap\L,\,
  j=0,\ldots,\n$, are marked $\CC_0,\CC_1,\ldots,\CC_\n$ and contain
  $N_0,N_1,\ldots, N_\n$ particles out of the infinitely many
  particles with positions and velocities denoted $\V X_0,\V
  X_1,\ldots,\V X_\n$, and $\dot{\V X}_0,\dot{\V X}_1,\ldots,$
  $\dot{\V X}_\n$, respecti\-vely, contained in $\O_j,\,j\ge0$ and
  considered in the $d=1,2$ cases. The second figure illustrates the
  special geometry considered for $d=3$ (as well as for $d=1,2$): here
  two thermostats, symbolized by the shaded regions, $\O_1,\O_2$
  occupy half-spaces adjacent to $\O_0$. } \*

A formal description of the model can be found in \cite{GP009,GP009a}.
The temperatures in the thermostats will be identified with their
average kinetic energy per particle. 

Considering external thermostats as correctly
representing the physics of the interaction of a system in contact
with external reservoirs has been introduced in \cite{WSE004}.
Their analysis was founded on the grounds of
\\
(1) identity, in the thermodynamic limit, 
of the evolution with and without thermostats
\\
(2) identity of the phase space contraction of the thermostatted
systems with the physical entropy production (up to a time derivative).

Here we follow their strategy.
To implement the physical requirement that the thermostats have well
defined temperatures and densities the initial data will be imagined
to be randomly chosen with a suitable distribution. If $\L$ is a ball
centered at the origin the space ${{\cal H}}(\L)$ will be the space of
the finite configurations $(\V X,\dot{\V X})$ with $\V
X\subset\cup_{j\ge0}\O_j\cap\L$.

Referring to Fig.1 denote by $K_j(\dot{\V X}_j)=\frac12\dot{\V X}_j^2$ the
kinetic energy of the particles in the $j$-th container (assuming the
particles mass $=1$), by $U_j(\V X_j)$ their potential energy (more
explicitly expressed in Eq.(\ref{e1.3}).

{\vskip2mm} \noindent {\bf Initial data:} {\it The probability
  distribution for the random choice of initial data will be,
  if $dx{\,{\buildrel def\over=}\,}\prod_{j=0}^\n\frac{ d\V
  X_j\,d\dot{\V X}_j}{N_j!}$, the
  distribution on ${\cal H}(\L)$
\be \m_{\L}(dx)= \frac{e^{-H_0(x)}}{Z(\L)} \prod_{j>0}\d(K_{j,\L}(x)-
\frac{d}{2\b_j}N_j)
\,dx\label{e1.1} \ee
with $H_0(x)=\sum_{j=0}^\n \b_j (K_j(\dot{\V X}_j)+ U_j(\V X_j))$, and
$N_j=\d_j |\L\cap\O_j|$ with $ \b_j \defi \frac1{k_BT_j}$ $>0,\d_j>0,
j>0$ ; the values $ \b_0=\frac1{k_BT_0}>0,\d_0>0$ will also be fixed
(but bear no particular physical meaning because the test system is
kept finite).}  {\vskip3mm}

Here $\Bd=(\d_0,\d_1,\ldots\d_\n)$ and ${\bf T}=(T_0,T_1,\ldots,
T_\n)$ are fixed {\it densities} and {\it temperatures}, $\L$ is a
ball centered at the origin and $Z$ is the normalization.

With the above choice of initial data the physical requirement that
the thermostats are in a configuration with densities and temperatures
assigned is realized at time $0$. But the probability distribution is
not invariant under time evolution: because the temperatures $\b_j$
are different and also because $H_0$ does not contain the interaction
between the system and the thermostats and because on the system will
be imgained to act nonconservative forces.

The equations of motion (see Fig.1) in $\L= \L_n$, with $\L_n$ being
the ball of radius $2^n r_\f$ with $r_\f$ a length unit that can be
chosen to be the interaction range (see below), will be different in
the {\it frictionless} thermostats and in the {\it isokinetic}
thermostats. They will differ by the value of a parameter $a=0,1$:

\beq
&m\ddot{\V X}_{0i}=-\partial_i U_0(\V X_0)-\sum_{j>0}
\partial_i U_{0,j}(\V X_0,\V X_j)+\BF_i(\V X_0)\nonumber\\
&m\ddot{\V X}_{ji}=-\partial_i U_j(\V X_j)-
\partial_i U_{0,j}(\V X_0,\V X_j)-a\a_j \dot{\V X}_{ji}
\label{e1.2}\eeq
where the first label, $j=0$ or $j=1,\ldots,\n$ respectively, refers to
the test system or to a thermostat, while the second indicates the
components of the coordinates of the points located in the
corresponding container and initially {\it in the regularization box $\L_n$}
(hence the labels $i$ in the subscripts $(j,i)$ have $N_j\,d $
values). Furthermore:
\*
\0(1) the $\BF(\V X_0)$ are, positional, {\it
nonconservative}, smooth stirring forces, possibly vanishing; 
\*\0(2) other forces are conservative and generated by a pair
potential $\f$, with range $r_\f$, which couples all pairs in the same
containers and all pairs of particles one of which is located in
$\O_0$ and the other in $\O_j$ ({\it i.e.} there is {\it no direct
  interaction} between the different thermostats).
\*\0(3) particles are repelled by the boundaries $\partial\O_j$ by a
conservative force of potential energy $\ps$, of range $r_\ps\ll
r_\f$, diverging at the walls. The potential energies will be $U_j(\V
X_j), \,j\ge0$, and $U_{0,j}(\V X_0,\V X_j)$: respectively denoting the
internal energies of the various systems and the potential energy of
interaction between the system and the thermostats.

\beq
&U_j(\V X)=\sum_{q\in\V X_j}\ps(q)+\sum_{(q,q')\in\V X_j,q\in\L}\f(q-q')\nn\\
&U_{0,j}(\V X_0,\V X_j)=\sum_{q\in\V X_0,q'\in\V X_j}\f(q-q')
\label{e1.3}\eeq
\*\0(4) in the case $a=0$ particles will be allowed to exit the
regions $\O_j\cap\L_n$ through the boundary of $\L_n$ while if $a=1$
they will be constrained to remain inside the container $\O_j\cap\L_n$
by an elastic reflection on $\partial\L_n$: this choice of the
boundary conditions is imposed mainly to make possible references to
\cite{GP009a}.
\*\0(5) the $\a_j$ are determined so that the solution to the
equations with $a=1$ {\it keeps exactly the same kinetic energy it has
at time $0$}, {\it i.e.} $\frac{d}2N_j k_B T_j$, for $j>0$, see
Eq.(\ref{e1.1}); this means

\be \a_j=\frac{Q_j-\dot U_j}{N_j \,d\,k_B T_j}\label{e1.4}\ee
where $Q_j\defi -\dot{\V X}_j\cdot\partial_{ \V X_j} U_{0,j}(\V X_0,\V
X_j)$, is the {\it heat} ceded by the system to the $j$-th thermostat.
\*\0(6) The potentials $\f,\ps$ have been chosen $j$--independent for
simplicity. The pair potential $\f$ will be supposed smooth, $\ge0$,
with finite range $r_\f>0$, setting $\f(0) $ $\defi\f_0>0$; the wall
potential $\ps$ will also be supposed smooth at a distance $r>0$ from
the walls, $\ge0$, with range $r_\ps$, and diverging proportionally to
$r^{-\a}$ as $r\to0$, for some $\a>0$.\*

\0{\bf Hypothesis:} {\it The initial state distributions $\m_{\L_n}$
are assumed to satisfy a {\it large deviations property} for the total
potential energy $U_{j,\L_n}(x)$; in the sense that for $C>0$ there is
$c>0$ such that:

\be \kern-2mm\m_{\L_n}\Big(\Big\{x\,\Big|\,
\big|\frac{U_{j,\L_n}(x)}{|\O_j\cap\L_n|}-u_j\big|
>\frac{ C\,\f_0}{N_j^\e}\Big\}\Big)
< e^{- c 2^{n(1-2\e) d}} \label{e1.5}\ee
for suitable $u_j\in\RRR,\e\in[0,1)$. This is a ``no phase
transition'' assumption (satisfied in the cluster expansion region of
the parameters $\V T,\Bl$ as discussed in \rm
\cite{GP009b,GM967}). Notice that the volume $|\O_j\cap\L_n|$ is
$O(2^{nd})$.}  \*

The analysis will be restricted to the geometries in  Fig.1
if $d=1,2$ and only in the second of Fig.1 if $d=3$.

\def\SEC{Time evolution}
\section{Time evolution}\label{sec2}
\iniz

Infinite systems are idealizations, not uncommon in statistical
mechanics, that must be considered as limiting cases of large, yet
finite, systems.  We therefore call {\it regularized} equations of
motion the Eq.(\ref{e1.2}) and denote $S^{(n,a)}_tx$, or also
$x^{(n,a)}(t)$, their solutions with data randomly chosen with the
distribution in Eq.(\ref{e1.2}).

Solutions exist (for all data in $\HH(\L)$, see the initial data
hypothesis in Sec.\ref{sec1}) by the standard existence
and uniqueness theorems for ordinary differential equations in the
case $a=0$.

The case $a=1$ involves elastic reflections on $\partial\L$ and a
proof of existence of the evolution requires (by adapting the analysis
in \cite{MPPP976}, see \cite[Appedix I]{GP009a}) obtaining exixtence
and uniqueness for all initial data {\it outside a set of
$\m_{\L_n}$-probability $0$}.

We shall discuss results that hold uniformly in the size $n$ provided
it is large enough.  The results can be conveniently formulated in
terms of quantities defined below.

Let $v_1\defi \sqrt{\frac{\f_0}m}$, $ |x_i-x'_i|\defi \frac{|\dot
  q_i-\dot q'_i|}{v_1}+\frac{|q_i-q'_i|}{r_\f}$, and

\beq 
&W(x;\x,R) \defi\frac1{\f_0}\sum_{q_i\in {\cal
B}(\x,R)}\Big(\frac{m\dot q_i^2}2
\nn\\
&+\frac12\sum_{j; j\ne
i}\f(q_i-q_j)+\ps(q_i)+\f_0 \Big)\label{e2.1}\\
&\NN_i(x)\defi \hbox{number of particles 
within $r_\f$ of $q_i$},\nn\eeq
Let $\log_+ z\defi\max\{1,\log_2|z|\}$, $g_\z(z)=(\log_+ z)^\z$,
$\r_n(q)$ = distance of $q$ from $\cup_j(\partial\O_j\cup \partial\L_n)$
and call ${\XX}_E\defi$ $\{x\,|\, {\cal E}(x)\le E\}$ with 

\be
\EE(x)\defi \sup_{\x} \sup_{R> g_{1/d}(x/r_\f)} \frac{W(x;\x,R) }{R^d}
\label{e2.2}\ee
The set $\XX_E$ has $\m_{\L_n}$--probability approaching $1$ as
$E\to\infty$ {\it uniformly in} $n$ (see, for instance,
\cite[Eq.(9.2)]{GP009a} choosing $c=E$ and $c'=1,\g(c)=c-c_0$):

\be \m_{\L_n}(\XX_E)\ge 1-\lis C e^{-\lis c E}\label{e2.3}\ee
for suitably chosen $\lis C,\lis c>0$.

Let $d\le3$, then given an initial datum $x$ the motion $S^{(n,a)}_tx$
  exists for $\m_{\L_n}$--almost all $x\in\HH_{1/d}$ and $\forall t\ge0$.
  Fixed {\it arbitrarily} an observation time $\Th<\infty$ the main
  result will be  \*

\0{\bf Theorem 1:} {\it If $x\in\HH(\L_n)\cap\XX_E$ and $0\le t\le\Th$
  there are $c=C(E,\Th)<\infty$, $c'=C'(E,\Th)>0, 1>\g>1/2$ such that, for
  particles which at time $0$ are in $\L_k$,
  i.e. $q_i(0)\in\L_k$. Consider, for $n>k$, the events

\beq
(1)&|\dot q_i^{(n,0)}(t)| \le\, c\, v_1 \,
k^{\frac12} ,\nn\\ 
(2)&\r_n(q_i^{(n,0)}(t))\ge c'\,r_\f
\,k^{-\frac1\a},\label{e2.4}\\
(3)&\NN_i(S^{(n,0)}_tx)\le c\,
k^{1/2},\nn\\
(4)&|(S^{(n,0)}_tx)_i-(S^{(n+1,0)}_tx)_i|\le
e^{-c'2^{n/2}} ,\ n>k,
\nn\\
(5)&\kern-2pt|(S^{(n,0)}_tx)_i-(S^{(n,1)}_tx)_i|\le
e^{-c'(\log n)^\g} ,\ k<(\log n)^\g.\nn\eeq
The events (1-4) are realized for all $x\in\XX_E$, while the
event (5) is realized with $\m_{\L_n}$-probability $\p_n$ and
$|\m_{\L_n}(\XX_E)-\p_n|\le c\, e^{-c'(\log n)^{2\g}}$.}

\* This means that, if $\L_n$ is large, motion of the particles close
to the test system is largely independent on the regularization size
$n$. And thermostatted motion and frictionless motions, near the test
system, are also very close. 

The uniformity in $n$ of the constants $c,c'$ is the really
interesting part of the statement.

Items (1-4) are proved in \cite[theorem 5, sec.6]{GP009a} for
$d=1,2,3$ (and for $d=1,2$, only, in \cite[theorem 7]{GP009} via a
different method). Therefore {\it the novelty in the present paper
will be the proof of item} (5): it will heavily rely on
\cite[Sec.VII]{GP009a}.

For later reference it is convenient to introduce a few more
notations.  

Let $C_\x$ the cube with side $r_\f$ centered at a point $\x$ in the
lattice $r_\f \ZZZ^d$. Let $\D$ be a subset of $\RRR^d$ and, using the
definitions in Eq.(\ref{e2.1}),

\beq
N_\D(x)\defi&\sum_{q\in \D} 1,\qquad V_\D\defi
\max_{q\in \D}\frac{|\dot
  q|}{v_1}\\
\|x\|_n\defi&\max_{\x\in\L_n} 
\frac{ \max(N_{C_\x}(x),{\e_{C_\x}(x)})}{g_{1/2}(\x/r_\f)},\label{e2.5}\eeq
where $\e_{C_\x}(x)\defi\sqrt{e_{C_\x}(x)}$, $e_C(x)=\max_{q\in C} (\frac12
\dot q^2+ \ps(q))$. 

\def\SEC{Thermostatted evolutions}
\section{Thermostatted evolutions}\label{sec3}
\iniz

Thermostatted evolution can now be studied by comparison with the
frictionless one.

The problem will be studied by restricting attention to a suitable
subset of the set $\XX_{E}\defi\big\{x\,|\, \EE(x)\le E\big\}$.

Consider the bands of points $\x$ at distance $\r_{0}(\x)$ within
$r_\f$ and $2r_\f$ from the boundary $\partial\O_0$ of $\O_0$:

\be \L_*\defi \{q: \r_{0}(q)\le r_\phi\},
\, \L_{**}\defi \{q: \r_{0}(q)\le 2r_\phi\}\label{e3.1} \ee
By items 1,2,3 theorem 1 in Sec.\ref{sec2} (which can be taken for
granted by the remark following theorem 1) there is $C_*>0$ (depending
on $E$) so that, for all $x\in \XX_E$ and with the notations
Eq.(\ref{e2.5}), for $n$ large enough:

\be\max_{t\le \Th} \,\max\{N_{\L_{**}}(S^{(n,0)}_{t}x),\,
V_{\L_{**}}(S^{(n,0)}_{t}x)\}< C_*\label{e3.2}\ee
Fixed $\g$ once and for all, arbitrarily with $\frac12<\g<1$, let
$\Si'$ be the set where are realized the events delimiting the {\it
stopping time} $T_n(x)$

\beq &T_{n}(x)\defi \big\{\max t : \,t\le
  \Th,\, \forall\, \tau< t, \label{e3.3}\kern2cm\\
&
\Big|\frac{U_{j,\L_n}(S^{(n,1)}_\t x)}
{|\O_j\cap\L_n|}-u_j\Big|<\frac{\f_0}{N_j^\e},
      \|S^{(n,1)}_\t x\|_{n} < (\log n)^{\g}\big\}.\nn\eeq
see  Eq.(\ref{e1.5}) and
Eq.(\ref{e2.5}) for notations.  Split $\XX_{E}=\AA_n\cup \BB_n$ with

\be\BB_n\defi\{x\in \XX_{E}: T_n(x)\le\Th\}:\label{e3.4}\ee
\*
 \noindent{\bf Theorem 2:} {\it There are positive constants $C,C',c$
 depending only on $E$ such that for all $n$ large enough:
\\
(1) if $t\le T_{n}(x)$, $S^{(n,0)}_tx$ and
 $S^{(n,1)}_tx$ are close in the sense that for $q_i(0)\in \L_{(\log
 n)^\g}$

\beq
&|q^{(n,1)}_i(t) -q^{(n,0)}_i(t)|\le \,C\, r_\f\, e^{-(\log n)^\g\,c },\nn\\
&|\dot q^{(n,1)}_i(t) -\dot q^{(n,0)}_i(t)|\le \,C\,v_1\, e^{-(\log
    n)^\g\,c} .
\label{e3.5}\eeq
Furthermore with the notations in Eq.(\ref{e3.1}),
(\ref{e3.2}), for $n$ large enough and for all ${t\le T_n(x)}$:

\be N_{\L_{*}}(S^{(n,1)}_{t}x)
\le C_*,\ V_{\L_{*}}(S^{(n,1)}_{t}x)\le  C_*+1
\label{e3.6}\ee
(2) the set $\BB_n$ has $\m_{\L_n}$--probability bounded  by

\be \mu_{\L_n}(\BB_n) \le \,C\,e^{-c (\log n)^{2\g}+C'}.\label{e3.7}\ee
}
\vskip2mm

\noindent{\it Remarks:} (1) The Eq.(\ref{e3.5}) together with the first
four items of theorem 1 imply Eq.(\ref{e3.6}).
\*
\0(2) A sequence of initial data $\{x\}=(x_1,x_2,\ldots)$ sampled
randomly and independently with the distributions $\m_{\L_n}$, {\it i.e.}
with the distribution $\m_0(d\{x\})\defi$ $\prod_{n=1}^\infty
\m_{\L_n}(dx_n)$, will consist of configurations $x_n\not\in\BB_n$ for all
$n$ large enough, by Borel-Cantelli's lemma and Eq.(\ref{e3.7}), because
$\g>1/2$.
\*\0(3) To prove item (1) and Eq.(\ref{e3.6}) we shall compare the evolutions
$x^{(n,1)}(t)$ with $ x^{(n,0)}(t)$, at same initial datum $x\in
\XX_E$ and $t\le T_{n}(x)$, the latter being the stopping time defined
in Eq.(\ref{e3.3}). 
\*

Two preliminary results are necessary, namely that there is $C>0$ so
that for all $n$ large enough the following holds.  \*

\noindent{\bf Lemma 1:} {\it Let $x\in\XX_E$, $t\le T_n(x)$, see
  Eq.(\ref{e3.3}), and $k\ge (\log n)^\g$, then
\beq
|\dot q^{(n,1)}_i(t)|&\le C\,v_1\, \big(k\,\log n)^\g,
\nn\\
|q^{(n,1)}_i(t)|&\le r_\f\,(2^k+C \, \big(k\,\log n)^\g).
\label{e3.8}\eeq
for $q_i(0)\in\L_k$ and $t\le\Th$.}
\*

This lemma is needed because, otherwise, the positions and speed at
time $t$ cannot be controlled in terms of the norms $\|x\|_n$ at time
$0$: since the particles move they must be followed (a ``Lagrangian''
viewpoint).
\*

A corollary of the above will be:
 \*

\0{\bf Lemma 2:} {\it Let $\NN$ and $\r$ be the maximal
  number of particles which at any given time $\le T_n(x)$ interact with a
  particle $q_i$ initially in $\L_{k+1}$ and, respectively, the minimal
  distance of a particle from the walls. Then

\be
\NN \le C\,(k \log n)^{d\g},\;\; \r \ge\, c\, (k \log n)^{-2(d\g+1) /\a}
  \label{e3.9}\ee
for all integers $k\in((\log n)^\g, 2(\log n)^\g)$.}  \*

\0{\it Remarks:} (1) $\a$ is the power which controls the divergence rate of the
wall potentials.
\\
(2) The proof of the lemmas is in \cite[Appendix L]{GP009a}.  \*

\0{\it Proof (of theorem 2):} The proof of the key ``entropy bound''
Eq.(\ref{e3.7}) is similar to the proof of the corresponding statement
in \cite{GP009a} and is reproduced in the Appendix A, because it
requires some changes with respect to the analysis in \cite{GP009a}
where the stopping time definition was based on the kinetic energy
rather then on the potential energy as in Eq.(\ref{e3.3}). In this
appendix we make essential use of the hypothesis at the end of Sec.1.

We shall now bound $\d_i(t,n) \defi$ $ |q_i^{(n,1)}(t)-
q_i^{(n,0)}(t)|$ again following \cite{GP009a}.  Let $f_i$ be the
acceleration of the particle $i$, with $q^{(n,a)}_i(t)\in \L_{k+1}$,
$t\le T_n(x)$, and $k\defi$ $ (\log_+ n)^\g$, due to the other particles
and to the walls.

If $q_i=q_i(0)\in \L_{k}$, $|f_i|\le C$ $\, (k\log n)^{\eta'}$,
$\eta'\defi (2d\,\g+1)\,(1+\frac 1\a)$ by Eq.(\ref{e3.9}). Consider the
two evolutions, for $a=0,1$ respectively,

\beq& q_i^{(n,a)}(t)=q_i(0)+\int_0^t \Big(e^{-\int_0^\t
a\a_j(x^{(n,a)}(s))ds}\dot q_i(0)\nn\\
&+\int_0^\t\,ds \,e^{-\int_s^\t
a\a_j(x^{(n,a)}(s'))ds'}\,f_i(x^{(n,a)}(s))\Big)\,d\t
\label{e3.10}\eeq
\0where the label $j$ on the coordinates (indicating the
container) is omitted and $f_i$ is the force acting on the selected
particle divided by its mass (for $j=0$ it includes the stirring
force).

Subtracting the Eq.(\ref{e3.10}) for $a=0$ and $a=1$ it follows that for
any $q_i\in \L_{k}$ (possibly close to the origin hence very far
from the boundary of $\L_k$ if $n$ is large, because $k=(\log_+ n)^\g$)

\beq &\d_i(t,n)\le C \, (k\log n)^{\eta'} 2^{-n\e d} \label{e3.11}
\\& + \Th \int_0^t |f_i(q^{(n,1)}(\t))-f_i(q^{(n,0)}(\t))|\,d\t.\nn\eeq
{\it provided} $|\int_{t_1}^{t_2} \a_j ds |$ is bounded proportionally
  to $2^{-n\e d}$ for $[t_1,t_2]\subset
[0,T_n(x)]$.
\*

From the definition Eq.(\ref{e1.4}) of $\a_j$ the bound of the
integral $|\int_{t_1}^{t_2} \a_j ds |$ with $[t_1,t_2]\subset
[0,T_n(x)]$ is split into a bound on $\int_0^{T_n(x)}
\b_j\frac{|Q_j|}{d N_j}dt$ and a bound on the ratio
$\b_j\frac{|U_{j}(S^{(n,1)}_{t_2}x)- U_{j}(S^{(n,1)}_{t_1}x)|}{d
N_j}$.  \*

The first bound can be derived from the inequality
$\|S^{(n,1)}_tx\|_n<(\log n)^\g$ ({\it i.e.} for $t\le T_n(x)$) and it
is $\le C $ $(k\log n)^\g 2^{-nd}$. This requires using lemmata 1,2
because $\|S_tx\|_n$ only gives information about the particle that at
time $0$ are close to the test system, while the $i$--particle might
be close to $\CC_0$ at time $t$ but not at time $0$.
\*

The second bound also follows from the definition of the stopping time,
which implies the validity of the inequality
$\frac{|U_{j}(S^{(n,1)}_{t_2}x)-U_{j}(S^{(n,1)}_{t_1}x)|}{d N_j}\le\,
C\, 2^{-n\e d}$. This also makes use of lemmata 1,2 (for the same
reason as above).  \*

Let $\ell$ be a non-negative integer, $k_\ell$ such that
\be 2^{k_\ell}=2^k +\ell \,C\, (k \log n)^{\g} \label{e3.12}\ee
and $u_{k_\ell}(t,n)$ the max of
$\d_i(t,n)$ over $|q_i|\le 2^{k_\ell}$.  

The difference in the accelerations is bounded, by lemma 2, by the
 maximum number $(k\log n)^{\g d}$ of the particles which can interact
 with $q_i(t,n)$ times $\max |\partial^2\f|$, plus a term proportional
 to $(k\log n)^{(2d\g+1)(1+2/\a)}$ due to the walls potential.  Then
 by Eq.(\ref{e3.11}) and writing $\eta''\defi (2\,d\,\g+1)(1+\frac
 2\a)$,

\beq&\frac{u_{k_\ell}(t,n)}{r_\f}\le \,C\, (k\log
  n)^{\eta'} 2^{-n\e d}\kern3cm\nn
\\& +C  (k\log n)^{\eta''}
\int_0^t\frac{ u_{k_{\ell+1}}(s)}{r_\f}\frac{ ds}\Theta.\label{e3.13}\eeq
for $\ell \le \ell^*=2^k/((k\log n)^\g C)$, the latter being the
largest $\ell$ such that $2^{k_\ell}\le 2^{k+1}$.  By Eq.(\ref{e3.13})

\beq u_{k}(t,n)\le& \,  e^{C\, (k\,\log n)^{\eta''}} C (k\log
  n)^{\eta'} 2^{-n\e d}
\nn\\&+ \frac{(C\, (k\,\log
    n)^{\eta''})^{\ell^*}} {\ell^*!}\; C\,(k \log n)^{\g}.\label{e3.14}\eeq

Thus for $n$ large enough $u_{k}(t,n)$ is bounded by the r.h.s.\ of
the first of Eq.(\ref{e3.5}); analogous argument shows that also the
velocity differences are bounded as in Eq.(\ref{e3.5}) which is thus
proved for all $t\le T_{n}(x)$.

It follows that, for $n> e^{k_0^{1/\g}}$ and $i$ fixed, given $q_i(0)$
with $|q_i(0)|/r_\f\le 2^{k_0}$ it is
$|q_i^{(n,1)}(t)-q^{(n,0)}_i(t)|<u_{(\log n)^\g}(t,n)\le C e^{-c(\log
n)^\g} $, {\it i.e.} for $n$ large $q^{(n,1)}_i(t)$ is closer than
$r_\f$ to $q^{(n,0)}_i(t)$. 

Remarking that we know ``everything'' about the Hamiltonian
motion we can use such knowledge by applying Eq.(\ref{e3.5}) to
particles which are initially within a distance $r_\f 2^{k_0}$ of the
origin, with $k_0$ fixed arbitrarily, for all large $n$.

Therefore the number of particles in $q^{(n,1)}_i(t)$ which are in
$\L_*$ is smaller than the number of particles of $q^{(n,0)}_i(t)$ in
$ \L_{**}$ which is bounded by $C_*$.  

An analogous argument for the velocities allows to complete the proof
of Eq.(\ref{e3.6}), given
the closeness of the positions and speeds of the motions with $a=0,1$,
see Eq.(\ref{e3.5}). Hence $T_{n}(x)\equiv \Th$ unless $x\in\BB_n$ and
the proof of theorem 3, hence of theorem 1, is complete.

\def\SEC{Appendix A: Entropy}
\section{Appendix A: Entropy}\label{sec4}
\iniz

Entropy production per unit time, in a configuration $x$, is naturally
defined in terms of $Q_j=-\dot{\V X}_j\cdot\partial_{ \V X_j}
U_{0,j}(\V X_0,\V X_j)$, interpreted as the {\it heat} ceded by the
system to the thermostats and it is given by:

\be \s_0(x)=\sum_{j>0}\b_j Q_j(x)\label{e4.1}\ee
In the isokinetic thermostat model, $a=1$, and {\it if the volumes in
phase space are measured by the distribution $\m_{\L_n}$}, a direct
computation shows that this quantity differs from the {\it contraction
rate} of the phase space volume by $\b_0( Q_0+\F(X_0)\cdot X_0)\equiv
\b_0(\dot K_0+\dot U_0)$ ($Q_0$ being $-\dot{\V
X}_0\cdot\partial_{X_0}U_{0,j}(\V X_0,\V X_j)$) and by a further
``small correction'' $-\frac{Q_j+\dot U_j}{dN_j}$, see also
\cite{Ga008d}. It is:

\be\s(x)=\s_0(x) -\sum_{j>0}\b_j \frac{\dot
U_j+Q_j}{d N_j} +\b_0(\dot K_0+\dot U_0).\label{e4.2}\ee

The Eq.(\ref{e3.6}) yields a bound $|\s_0(x)|\le C$,
$\frac{|Q_j|}{N_j}\le C 2^{-nd}\le C$ and also the last term in
Eq.(\ref{e4.2}) gives a similar bound (because the number of particles
in $\O_0$ is bounded in terms of $E$ and $\dot K_0+\dot U_0$ equals
$Q_0+\F(\V X_0)\cdot \dot{\V X}_0$ is bounded by Eq.(\ref{e3.6}) and,
via the comparison with the frictionless motion in Eq.(\ref{e3.5}) and
the first four items in theorem 1).

Furthermore by the definition of the
stopping time it is $\int_{t_1}^{t_2}\frac{\dot U_j}{N_j}dt\le C
2^{-n\e d}$: so that there is $C'$ and

\be \int_{t_1}^{t_2} |\s(S^{(n,1)}_tx)| dt \le C',\qquad
\forall\,t_1\le t_2\le T_n(x)\label{e4.3}\ee
Writing $k_\xi$ for the smallest integer $\ge (\log n)^\g
g_\g(\xi/r_\phi)$ (here $g_\g$ is chosen instead of the natural
$g_{1/2}$ in order to simplify the formulae: recall that by
defiinition $\g>1/2$), then codimension $1$ surface containing $\Si'$
splits into an union over $\xi \in \Lambda_n\cap r_\f\ZZZ^d$ of the
union of $\mathcal S^1_\xi\cup \mathcal S^2_\x\cup{\cal S}^3$, where

\beq \mathcal
S^1_\xi=&\{y\in \Si':|y\cap C_\xi|=k_\xi, |y\cap
\partial C_\xi|=1\}\nn\\
\mathcal S^2_\xi=&\{y\in \Si':y\cap C_\xi\ni
(q,\dot q), \e(q,\dot q)=\wt \e_\x\}\cr
{\cal S}_\pm^3=& \{y\in \Si':
\frac{U_{j,\L_n}(x)}
{
|V_n|}-u_j=\pm\frac{\f_0}{2^{n \e d}}
\}
\label{e4.4} \eeq
if $\wt\e_\xi\defi \big((\log n)^\g g_{\g}(\xi/r_\phi)\big)^2$ and
$V_n\defi\O_j\cap\L_n$.
\*

{\it By the assumption Eq.(\ref{e1.5}) it is not restrictive to suppose that
at $t=0$ it is $\big|\frac{U_{j,\L_n}(x)} {
|V_n|}-u_j\big|<\frac12 \frac{\f_0}{2^{n \e d}} \}$.}
\*

Abridge $\D_{j,n}\defi \frac{U_{j,\L_n}(x)}
{|V_n|}-u_j$ and consider first the case of ${\cal S}_+^3$.
Let $D\subset {\cal S}_+^3$ be the set of the $x$ which satisfy
$\D_{j,n}(x)=\f_02^{-n \e d}$ for a given $j>0$ while
$\D_{j',n}U(x)<\f_02^{-n \e d}$ for $j'>0,\,j'\ne j$.

Then the probability $\m_{\L_n}(D)$
can be bounded, see Eq.(\ref{e3.2}), by

\beq & e^{C'}\int \m_{\L_n}(dx)\,
\Th\,\d(\D_{j,n}-G)\,|\widehat G|\,\ch(N\le \r 2^{nd})\label{e4.5}\eeq
where $G$ is any value between $\frac12\frac{\f_0}{|V_n|}$ and
$\frac{\f_0}{|V_n|}$ and $\widehat G(x)$ is the derivative $\dot
\D_{j,\L}$ {\it i.e.}, from the equations of motion,

\be\widehat G(x)\defi \frac1{|V_n|}(\sum_{q,q'}
(\dot q-\dot q') \partial_q
\f(q-q')+\sum_q \dot q \partial_q\ps(q))\label{e4.6}\ee
and $e^{C'}$ takes into account the entropy estimate {\it i.e.} the
bound Eq.(\ref{e4.3}) of the non-invariance of $\m_{\L_n}$.  

Averaging the bound in Eq.(\ref{e4.6}) over $G$ between $\frac12
\frac{\f_0}{|V_n|}$ and $\frac{\f_0}{|V_n|}$, our bound is

\be e^{C'}|\frac{|V_n|}{\f_0} \int \m_{\L_n}(dx)|\widehat
G(x)|\ch_{x}\label{e4.7}\ee
where $\ch_x$ is the characteristic function of the set where
$|\D_{j,n}(x)|\in [\frac12 \frac{\f_0}{|V_n|},\frac{\f_0}{|V_n|}]$.
By Schwartz's inequality this is

\be\le e^{C'}\frac{|V_n|}{\f_0} (\m_{\L_n}(\widehat G^2)^{1/2}
\m_{\L_n}(\ch_x)^{1/2})\label{e4.8}\ee
By superstability and the integral can be bounded above by $C$ while
the second integral estimates the square root of a probability of a
lage deviation and is therefore bounded by by $e^{-c 2^{n
(1-2\e)d}}$. Hence the contribution of $\CC_3$ to the bound on
$\m_{\L_n}(\BB)$ is (amply) bounded by the {\it r.h.s.} of
Eq.(\ref{e3.7}).

Similarly, the surface areas $\m_{\L_n,\Si'}({\cal S}^1_\x)$ and
$\m_{\L_n,\Si'}({\cal S}^2_\x)$ induced by $\m_{\L_n}$ on ${\cal
S}^1,{\cal S}^2$ induced by $\m_0$ are bounded by

\be \m_{\L_n,\Si'}({\cal S}^i_\x)\le C \sqrt{n} e^{-c
[(\log n)^\g g_{\g}(\x/r_\f)]^2 }, \label{e4.9}\ee

\0(for suitable $C,c$, functions of $E$), also summable in $n$: this
bound can be taken from Appendix J in \cite{GP009a} where it is
derived in detail in the text following Eq.(9.41).


\bibliographystyle{unsrt}

\begin{thebibliography}{1}

\bibitem{GP009}
G.~Gallavotti and E.~Presutti.
\newblock Thermodynamic limit of isoenergetic and hamiltonian thermostats.
\newblock {\em arXiv:}, 0903.3316:1--9, 2009.

\bibitem{GP009a}
G.~Gallavotti and E.~Presutti.
\newblock Nonequilibrium, thermostats and thermodynamic limit.
\newblock {\em Journal of Mathematical Physics, arXiv:},
  51: 015202 (+32), 2010.

\bibitem{WSE004}
S.R. Williams, D.J. Searles, and D.J. Evans.
\newblock Independence of the transient fluctuation theorem to thermostatting
  details.
\newblock {\em Physical Review E}, 70:066113 (+6), 2004.

\bibitem{GP009b}
G.~Gallavotti and E.~Presutti.
\newblock Fritionless thermostats and intensive constants of motion.
\newblock {\em in print in Journal of Statistical Physics}, 2010.

\bibitem{GM967}
G.~Gallavotti and S.~Miracle-Sol\'e.
\newblock Statistical mechanics of lattice systems.
\newblock {\em Communications in Mathematical Physics}, 5:317--323, 1967.

\bibitem{MPPP976}
C.~Marchioro, A.~Pellegrinotti, E.~Presutti, and M.~Pulvirenti.
\newblock On the dynamics of particles in a bounded region: A measure
  theoretical approach.
\newblock {\em Journal of Mathematical Physics}, 17:647--652, 1976.

\bibitem{Ga008d}
G.~Gallavotti.
\newblock On thermostats: {I}sokinetic or {H}amiltonian? finite or infinite?
\newblock {\em Chaos}, 19:013101 (+7), 2008.

\end{thebibliography}
\small %

\end{document}